\documentclass{article}
\usepackage[margin=1in]{geometry}
\usepackage[utf8]{inputenc}
\usepackage{amsfonts}       
\usepackage{nicefrac}       
\usepackage{microtype}      
\usepackage[export]{adjustbox} 
\usepackage{amsmath}
\usepackage[colorinlistoftodos]{todonotes}
\usepackage{bm}
\usepackage{breqn}
\usepackage{physics}
\usepackage{enumitem}
\usepackage{bbold}
\usepackage{epigraph}

\usepackage{csquotes}
\usepackage{hyperref}       
\usepackage{url}            
\usepackage{booktabs}       
\usepackage{amsfonts}       
\usepackage{nicefrac}       
\usepackage{microtype}      
\usepackage{color}
\usepackage{scalerel}
\usepackage[toc,page]{appendix}
\usepackage{tabularx,ragged2e,booktabs,caption}
\usepackage[font=small,labelfont=bf]{caption}
\usepackage{blindtext}
\usepackage{amssymb}
\usepackage{mathtools}
\usepackage{dsfont}
\usepackage{amsmath}
\allowdisplaybreaks
\usepackage{amsfonts}
\usepackage{amsthm}
\usepackage{float}
\usepackage{graphicx}
\usepackage{subcaption}
\graphicspath{ {images/} }
\usepackage{letltxmacro}%
\usepackage{thmtools,thm-restate}
\usepackage{relsize}
\usepackage{algorithm,algpseudocode}
\usepackage{titlesec}
\usepackage{multirow}

\title{Two Burning Questions on COVID-19:
\\ {\smaller Did shutting down the economy help? Can we (partially) reopen the economy without risking the second wave?}}
\author{Anish Agarwal, Abdullah Alomar, Arnab Sarker, Devavrat Shah, Dennis Shen, Cindy Yang\thanks{Contact Information: anish90@mit.edu, aalomar@mit.edu, arnabs@mit.edu, devavrat@mit.edu, deshen@mit.edu, cxy99@mit.edu}}
\date{MIT}

\newcommand{\Ic}{\mathcal{I}}
\newcommand{\hlambda}{\widehat{\lambda}}
\newcommand{\Rb}{\mathbb{R}}
\newcommand{\hy}{\widehat{y}}
\newcommand{\ty}{\tilde{y}}
\newcommand{\bY}{\boldsymbol{Y}}
\newcommand{\btY}{\boldsymbol{\tilde{Y}}}

\begin{document}

\maketitle

\begin{abstract}
As we reach the apex of the COVID-19 pandemic across the globe, the most pressing question facing us all is: can we, even partially, reopen the economy without risking the second wave? Towards answering this question, we first need to understand if shutting down the economy helped. And second if it did, is it possible to achieve similar gains in the war against the pandemic while partially opening up the economy? 
To do so, it is critical to understand the effects of the various interventions that can be put into place and their corresponding health and economic implications.
Since many possible interventions exist, the key challenge facing policy makers is understanding the potential trade-offs between them, and choosing the particular set of interventions that works best for their circumstance 
-- for example, what would be the effect on COVID-19 related deaths if the U.S. had decreased mobility by 20\% instead of 60\% versus if Italy had done the same?

In this memo, we provide an overview of Synthetic Interventions \cite{SI} (a natural generalization of the widely applied Synthetic Control method \cite{abadie_survey}), which is a data-driven and statistically principled method to perform {\em what-if} scenario planning, i.e., for policy makers to understand the trade-offs between different interventions before having to actually enact them. 
In essence, the Synthetic Interventions method leverages information from different interventions that have already been enacted across the world and fits it to a policy maker's setting of interest 
-- for instance, to estimate the effect of mobility-restricting interventions on the U.S., we use daily death data from countries that enforced severe mobility restrictions to create a ``synthetic low mobility U.S.'' and predict the {\em counterfactual} trajectory of the U.S. if it had indeed applied a similar intervention.

We use the Synthetic Interventions method towards answering the two questions laid out above via a case study of the effectiveness of different interventions in mitigating the COVID-19 pandemic at the international level.
Encouragingly, we find that lifting severe mobility restrictions and only retaining moderate mobility restrictions (at retail and transit locations), seems to effectively {\em flatten the curve}. 
We hope that this will provide useful guidance for policy makers as they weigh the trade-offs between the safety of the population, strain on the healthcare system, and impact on the economy.


\end{abstract}

\section{Importance of What-If Scenario Planning}

\paragraph{Challenge of Choosing Between Interventions.}
It is clear that the COVID-19 pandemic has led to an unprecedented disruption of modern society at a global scale.
What is much less clear, however, is the effect that various interventions that have been put into place have had on health and economic outcomes. 
For example, perhaps a 30\% and 60\% clampdown in mobility have similar societal health outcomes, yet vastly different implications for the number of people who cannot go to work or file for unemployment.
Having a clear understanding of the trade-offs between these interventions is crucial in charting a path forward on how to open up various sectors of society. 
A key challenge is that policy makers do not have the luxury of actually enacting a variety of interventions and seeing which has the optimal outcome.
In fact, at a societal level, this is simply infeasible.
Arguably, an even bigger challenge is that the COVID-19 pandemic, and the resulting policy choices ahead of us, are unprecedented in scale. Thus, it is difficult to reliably apply lessons from previous pandemics (e.g., SARS, H1N1).
This is only further exacerbated when taking into the account the vastly different economic, cultural, and societal factors that make each town/city/state/country unique.
 
%
%
\paragraph{Synthetic Interventions.}
In this memo, we introduce Synthetic Interventions \cite{SI}, a method towards solving the problem laid out above. 
Specifically, Synthetic Interventions provides a data-driven and statistically principled way to perform {\em what-if} scenario planning, i.e., for policy makers to understand the trade-offs between different interventions before having to actually enact them. 
In essence, the Synthetic Interventions method leverages information from different interventions that have already been enacted across the world and fits it to a policy maker's setting of interest 
-- for example, to estimate the effect of mobility-restricting interventions on the U.S., we use daily death data from countries that enforced severe mobility restrictions to create a ``synthetic low-mobility U.S.'' and predict the {\em counterfactual} trajectory of the U.S. if it had indeed applied a similar intervention.
We highlight a few desirable attributes of this methodology. 
\begin{itemize}[leftmargin=*]
    \item {\em Personalized Predictions} -- the method takes into account the heterogeneity of the particular (geographical) purview of a policy-maker. For example, the method would provides different predictions for the effects of a 40\% drop in mobility for the U.S. vs. India vs. Italy etc. based on the particulars of that country.
    \item {\em Simplicity \& Interpretability} -- the method relies on building ``synthetic'' versions of each geographical location under different interventions by simply using a weighted combination of geographical locations that did indeed enact such an intervention. Thus, Synthetic Interventions requires relatively little hyper-parameter tuning. 
    We hope that the simplicity, interpretability, and robustness (yet surprising accuracy) of the method will encourage policy makers to apply Synthetic Interventions without fear of over-fitting to the idiosyncracies of their data. 
    %
    \item {\em Low Data Requirement} -- the method produces accurate forecasts using only (i) a few ``donor'' regions, i.e., the method only requires a small number of regions, approximately 5-10, to have gone through the intervention of interest; (ii) measurements over a small number of time periods, approximately 10-30 days (this can be viewed as the amount of training data); (iii) data corresponding to the metric of interest from each geographical location (e.g., daily national death rates), i.e., it does not require additional covariate information about each location; however, if auxiliary information is available, then the Synthetic Interventions method can naturally incorporate these data points in the model learning procedure as well.
\end{itemize}

%
%
\paragraph{COVID-19 Case-Study using Synthetic Interventions.}
In the next section, we detail the use of Synthetic Interventions to evaluate the effect of various mobility restriction interventions on health outcomes.
We give attention to the data requirements of the method, some important algorithmic details, and how to evaluate whether the method is producing accurate forecasts.
We hope this memo serves as a template for policy makers to make statistically informed decisions on the variety of weighty policy choices ahead of them.

\section{Synthetic Interventions: A Practitioner's Guide}
Recall that the aim of Synthetic Interventions is to predict the counterfactual trajectories of a region under a variety of interventions of interest.
Below, we explain how policy-makers can use the Synthetic Interventions method to help guide them in choosing between interventions.
For concreteness, we use the case study of applying Synthetic Interventions to understand the effect of mobility restriction policies on the number of COVID-19 related deaths at a country level, as a running example.

\subsection{Modeling Decisions \& Data Requirements}
We begin by listing some of the key modeling decisions and data requirements for the Synthetic Interventions method to produce reliable counterfactual predictions. 

%
\paragraph{Choosing Metric of Interest.}
First, a practitioner must decide upon his or her relevant metric of interest.
%
Some examples include the number of (i) COVID-19 related deaths, (ii) COVID-19 related cases, (iii) ICU beds utilized, and (iv) unemployment claim filings.
%
The counterfactual predictions produced by the Synthetic Interventions method will be with respect to the particular metric chosen.  

\paragraph{Choosing Interventions of Interest.}
Secondly, a practitioner must define a set of interventions that are of interest to the policy maker. 
%
For example, in the case of studying the effect of mobility, the interventions could be a (i) less than $10\%$ drop in mobility, (ii) $10-40\%$ drop in mobility, and (iii) greater than $40\%$ drop in mobility compared to the national baseline.
The counterfactual predictions produced by the Synthetic Interventions method will be for the particular interventions chosen. 

\paragraph{Choosing Pre- and Post-Intervention Time Periods.}
The third decision that needs to be made is defining the pre- and post-intervention periods, i.e.,
the data that is fed into the algorithm must be divided into two segments: 
\begin{itemize}[leftmargin=*]
    \item {\em Pre-Intervention} -- 
        this is the time period when all geographic locations have not yet enacted an intervention (or, more generally, enacted a common intervention); 
        in our context, the pre-intervention period is defined by the time horizon in which all countries are operating as normal, prior to any COVID-19 related outbreaks (i.e., none of the countries have implemented any mobility-restricting policies). 
        %
        Properly defining this time period is crucial as the algorithm strictly requires that the countries of interest are all operating under a common setting. 
        %
        %
    \item {\em Post-Intervention} -- 
        the post-intervention period should be defined as the time period when we see the effect of the interventions that each geographic location enacted. 
        Importantly, we note that the pre- and post-intervention periods do not have to be contiguous, chronological time instances.
        %
\end{itemize}

\paragraph{Categorizing Geographical Locations by Intervention Received.}
The fourth decision that needs to be made is categorizing the various geographic locations by the interventions they enact, which are observed in the post-intervention period. 
In our context, this would be categorizing the countries by the average drop in mobility compared to the national baseline in the pre-intervention period.

\subsection{Input, Output of Synthetic Interventions Method}
\paragraph{Algorithm Overview.}
The method has two major steps:
\begin{itemize}[leftmargin=*]
    \item {\em Synthetic Target Unit Construction} -- construct a ``synthetic'' model of the {\em target} region (i.e., the region we are interested in producing counterfactuals for) as a weighted combination of other {\em donor} regions (i.e., regions that did indeed enact the intervention of interest).
    \item {\em Counterfactual Prediction} -- use the model of the ``synthetic'' target unit, along with the observations of our donor units under our intervention of interest, to create the target unit's counterfactual trajectory in this parallel setting.
\end{itemize}

\paragraph{Data Input.}
The data input into the Synthetic Interventions algorithm is divided into two parts: 
\begin{itemize}[leftmargin=*]
    \item {\em Pre-Intervention Target and Donor Group Data} -- for the pre-intervention period, we require time series data for the metric of interest (e.g., daily death rates) associated with (i) the target region; %
    (ii) and, for every intervention of interest, the corresponding donor regions (i.e., a donor group) that receive that intervention during the post-intervention period. 
    As mentioned earlier, it is important that 
    all regions (target and donors)
    do not receive any intervention during the pre-intervention period. 
    \item {\em Post-Intervention Donor Data} -- 
    for the post-intervention period, we require time series data for each donor region under the interventions of interest.
    %
\end{itemize}

\paragraph{Data Output.}
The result of applying the Synthetic Interventions method on the data input described above would be {\em a counterfactual trajectory} of the target unit in the post-intervention time period under {\em every intervention of interest}. In other words, for each donor subgroup (which receives a unique intervention), one counterfactual trajectory for the target unit will be produced.

\paragraph{Mathematical Formulation.}
Please refer to Section \ref{sec:math_formulation} for a formal description of the Synthetic Interventions method.

\section{COVID-19 Case Study} 
 
\subsection{Modeling Decisions Made to Reliably Use Synthetic Interventions}\label{sec:case_study_modeling}

\paragraph{Choosing Metric of Interest - Daily Death Counts.}
Although there are a myriad of ways to measure a region's health and economic outcomes, some metrics such as the number of cases are considered ``noisier'' due to the inconsistencies of testing across regions, and others such as the number of daily unemployment filings are sparse. 
As a result, given its greater reliability and availability, daily death count reports are used as our outcome variable of interest.
Thus, we will apply the Synthetic Interventions method to predict how different interventions affect the number of COVID-19 related deaths within a region. 

\paragraph{Choosing Interventions of Interest - Daily Mobility Rate.} 
Each country has implemented a variety of interventions simultaneously to combat the spread of COVID-19. This makes it difficult to analyze any particular intervention (e.g., stay-at-home orders vs. schools shutting down) in isolation. 
However, if we consider the causal chain of events, we observe that all of these interventions are directed towards restricting how individuals move and interact, which then affects the spread and, ultimately, deaths associated with COVID-19. 
In light of this, we adopt mobility as our notion of intervention, and investigate how a country's drop in mobility level translates to the number of potential COVID-19 related deaths\footnote{
To resolve any potential ambiguities, we will henceforth refer to change in the level of mobility within a country as an intervention, and refer to orders such as the closing of schools or the enforcing of individuals to stay at home as a policy (which eventually translates to an intervention). 
}.
To that end, we use Google's excellent mobility reports \cite{google_mobility} to define three distinct, mutually exclusive interventions as follows:
\begin{itemize}
    \item[(a)] {\em Low Mobility Restricting Intervention} -- reduction in mobility is below $10\%$ compared to national baseline from January 2020.
    
    \item[(b)] {\em Moderate Mobility Restricting Intervention} -- reduction in mobility is between $10\%$ to $40\%$ compared to national baseline from January 2020.
    
    \item[(c)] {\em Severe Mobility Restricting Intervention} -- reduction in mobility is greater than $40\%$ compared to national baseline from January 2020.
\end{itemize}

\paragraph{Choosing Pre- and Post-Intervention Time Periods - Number of Deaths in the Country.} 
As previously mentioned, it is crucial to have a well-defined pre- and post-intervention period, where all regions during the pre-intervention period operate under a no intervention setting, and the effects of their enacted interventions are only observed during the post-intervention period.
Towards this, an assumption we rely on is that the enactment of mobility lockdown in a country is a function of the number of deaths in the country instead of a particular calendar date.
This is indeed verified in a data-driven manner using the Google mobility reports -- we see that 20 days prior to 80 deaths in a country (and any time before that), essentially no country enacted a mobility lockdown.
Thus, we align the daily death rate trajectories, and correspondingly the pre- and post-intervention periods, as before and after the country has 80 deaths.
In other words, we refer to the day a country hits 80 deaths as ``Day 0'', and the pre- and post-intervention periods then refer to the days before and after Day 0, respectively.

\paragraph{Categorizing Countries by Intervention Received -- Average (Lagged) Mobility Score.}
Studies have shown that there is a median lag of 20 days from the onset of infection to the day of death (e.g., see \cite{lag_study}).
Hence, a country's daily death count is a function of the region's infection levels, which are in turn affected by mobility levels, from roughly 20 days prior.
Thus, to compute the effect of a mobility intervention from Day 0 onwards, we compute a country's mobility score (i.e., intervention level) from Day -20 to Day -1.
Given that the mobility score in \cite{google_mobility} is changing every day, we take the average mobility score of a country from Day -20 to Day -1 and then bucket it into the three intervention groups defined above\footnote{Moving forward, we aim to apply more sophisticated clustering methods such as $K$-means clustering.}.

\subsection{Empirical Results -- Looking Back: What Could Have Happened?} \label{sec:looking_back} 
Here, we apply the Synthetic Interventions method to produce the counterfactual predictions of the daily death counts for all countries in the Google mobility data set under the three possible interventions detailed in Section \ref{sec:case_study_modeling}.

\paragraph{Validating Synthetic Interventions.}
Our aim is to validate that the method is able to produce accurate counterfactual predictions across all three interventions. 
To that end, for every intervention (mobility restriction level), we display the counterfactual predictions associated with two representative countries that enacted that intervention\footnote{Similar results hold generally across all countries. Please contact authors for any specific country of interest.}.
In the low mobility restricting regime, we show results for the United States and the United Kingdom in Figures \ref{fig:us} and \ref{fig:uk}, respectively, in Appendix \ref{sec:fig_pics1}. 
Pleasingly, we observe that the output produced by Synthetic Interventions is able to accurately recreate the observed death rates in the post-intervention period. 
We note that the model is learned during the pre-intervention period (Day -20 to Day -1), and the dashed lines in Days 0 - 15 are the predicted values under all possible mobility restriction levels (while solid lines are the true values). 

\vspace{2mm}
\noindent
Similarly, in the moderate and severe mobility restricting regimes, we display the results for Turkey, Sweden, India, and Romania in Figures \ref{fig:turkey}, \ref{fig:sweden}, \ref{fig:india}, and \ref{fig:romania}, respectively, in Appendix \ref{sec:fig_pics1}. 
Again, we observe that the counterfactual predictions produced from the Synthetic Interventions method is able to closely match the observed death rates under all different interventions, i.e., mobility restrictions. 
Importantly, we note that in all of these models, the Synthetic interventions built for the target country is learned in the pre-intervention period, when no intervention has yet occurred. 
Still, we are able to transfer the learnt model into an interventional setting, i.e., when the interventions take effect within the donor countries. 

\vspace{2mm}
\noindent
For each of the countries listed above, we display their top four donor countries (under each intervention) that most closely resemble them. These are shown in Figures \ref{fig:donor_us}, \ref{fig:donor_uk}, \ref{fig:donor_turkey}, \ref{fig:donor_sweden}, \ref{fig:donor_india}, and \ref{fig:donor_romania} in Appendix \ref{sec:fig_pics1}.  

\paragraph{Key Takeaway from Counterfactual Predictions.}
A stark conclusion that is immediately apparent from the figures above is that uniformly across all countries, there is a significant drop in the number of deaths with a even a ``small'' drop in mobility (i.e, a 10-40\% drop in mobility compared to the national baseline).
After this point, gains by further restricting mobility seem to be diminishing.
An ongoing line of research is to study how these various restrictions in mobility affect economic outcomes of a country.
This will ideally allow policy-makers to precisely quantify the trade-off between various mobility restricting interventions.

\subsection{Empirical Results -- Looking Ahead: What Can Happen?} \label{sec:looking_forward}

\paragraph{What Can Happen if Mobility Restrictions are Lifted?} 
Using the output of Synthetic Interventions, a natural next question that arises is estimating the peak number of daily deaths if mobility restrictions are lifted as of today (April 23rd, 2020).
To do so, we input the counterfactual estimates of the Synthetic Interventions method into an exponential curve-fitting algorithm.  
Specifically, for each country, we fit an exponential function to its current trajectory and each intervention level that is less restrictive than its current intervention level. 
We display the corresponding figures in Appendix \ref{sec:fig_pics2}. 

\paragraph{Key Takeaways.}
In a similar vein to the previous section, we see from the figures in Appendix \ref{sec:fig_pics2}, even small drops in mobility can significantly decrease the peak number of daily deaths in a country. 
Given that the peak number of daily deaths is the key metric in understanding how strained each nation's medical system will be, we see that moderate mobility restriction measures can lead to a significant ``flatenning'' of the curve. 
%


\section{Synthetic Interventions - Formal Description} \label{sec:math_formulation}

\paragraph{Setup.}
We consider the setting where there are $N$ units and $D$ interventions, and we observe the outcome variable of interest (e.g., death count) associated with each unit across $T$ time periods. 
Throughout, we let $n, t$, and $d$ be the indices associated with unit $n$, time $t$, and intervention $d$, respectively. 
Further, we denote $T_0 < T$ as the intervention point (e.g., Day 0), which partitions our time horizon into two distinct segments: 
the {\em pre-intervention} period ($t \le T_0$), where all units are assumed to be in a common setting (e.g., in the case study above, all countries operate as usual prior to any COVID-19 related outbreaks), 
and a {\em post-intervention} period ($t > T_0$), where each unit may receive one of $D$ interventions (e.g., a country goes into a low/moderate/severe mobility lockdown). 

\paragraph{Potential Outcome Notation.}
We let $y^{(d)}_{nt}$ denote the potential outcome for unit $n$ at time $t$ under intervention $d$. 
Without loss of generality, we will let $d=1$ represent the absence of any intervention or {\em null}-intervention (our assumed common pre-intervention setting).
Finally, we let $\Ic^{(d)}$ represent the set of units that actually receive intervention $d$ (e.g., mobility level $d$) in the post-intervention period, and $N^{(d)} = \abs{\Ic^{(d)}}$ as the number of units within partition $\Ic^{(d)}$.
%

\paragraph{Input.}
%
For each unit, we have access to a time series of $T$ observations\footnote{It is not necessary that the observations are indexed by time, though it is the most common way datasets are structured. Rather, we require $T$ (pre- and post-intervention) measurements for each of the $N$ units.} 
according to some outcome variable of interest. 
We can represent the available data (i.e., the dataset that is fed into the synthetic interventions method) as an $N \times T$ matrix $\bY = [y_{nt}] \in \Rb^{N \times T}$, with $(n,t)$-th entry  expressed as follows:
\begin{align*}
    y_{nt} &= \begin{cases}
        &y^{(1)}_{nt}, \text{ for all } t \le T_0
        \\ &y^{(d)}_{nt} \cdot \mathbb{1}(n \in  \Ic^{(d)}) + \star \cdot \mathbb{1}(n \notin  \Ic^{(d)}), \text{ for all } t > T_0, d \in [D]
        \\ &\star, \text{ otherwise,}
    \end{cases}
\end{align*}
where $\star$ indicates a missing value and $\mathbb{1}$ denotes the indicator function. 
That is, during the pre-intervention period, the observed outcomes associated with all units are associated with the null-intervention. During the post-intervention, however, we observe the outcome variable associated with each unit based on the intervention it received. 
%

\paragraph{Output.} 
The output of the Synthetic Interventions algorithm is simply an augmentation of the observed data $\bY$ under all possible interventions. 
That is, the Synthetic Interventions algorithm predicts the counterfactual trajectory of the outcome variable under all $D$ interventions for every unit. 
Hence, for each unit $n$, synthetic interventions produces $\hy^{(d)}_{nt}$ for all $(t, d)$, where $t > T_0$ and $d \in [D]$. 
If the counterfactual predictions are accurate and reliable, then a practitioner now has insights as to how each intervention affects every unit across time, and can reason about the trade-offs between interventions. 
Importantly, we note that these trajectories are unit specific, i.e., the counterfactual trajectories produced will be unique to each unit.

\subsection{Algorithm}
For notational simplicity (and without loss of generality), we temporarily consider unit $n=1$ as our {\em target} unit of interest, which operates under intervention $d^*$ in the post-intervention period; 
we will refer to the remaining units as our {\em donor} units. In what follows, we describe how to predict the counterfactual outcome variables for our target unit if it was operating under a different intervention $d \neq d^*$ instead. 
The same procedure can then be applied towards all remaining units. 

\vspace{1mm}
\noindent
For convenience, we will denote $\bY^{(d, \text{pre})} = [y^{(1)}_{nt}]_{n \in \Ic^{(d)}, t \le T_0}$ and 
 $\bY^{(d, \text{post})} = [y^{(d)}_{nt}]_{n \in \Ic^{(d)}, t > T_0}$ as the pre- and post-intervention matrices, respectively, associated with the donor units who received intervention $d$. 
Let their singular value decompositions (SVDs) be denoted as: 
 \begin{align*}
     \bY^{(d, \text{pre})} &= \sum_{i=1}^{N^{(d)} \wedge T_0} \sigma_i u_i v_i^T
     \qquad \text{and} \qquad \bY^{(d, \text{post})} = \sum_{i=1}^{N^{(d)} \wedge (T-T_0)} s_i \mu_i \nu_i^T.  
 \end{align*}

\begin{enumerate}
    \item {\em Data de-noising, regularizing:} perform Singular Value Thresholding (SVT) on $\bY^{(d, \text{pre})}$ and $\bY^{(d, \text{post})}$ to produce $\btY^{(d, \text{pre})} = [\ty^{(1)}_{nt}]_{n \in \Ic^{(d)}, t \le T_0}$ and 
 $\btY^{(d, \text{post})} = [\ty^{(d)}_{nt}]_{n \in \Ic^{(d)}, t > T_0}$, denoted as: 
    \begin{align} \label{eq:denoise}
        \btY^{(d, \text{pre})} = \sum_{i=1}^{r^{(d, \text{pre})}} \sigma_i u_i v_i^T
        \qquad \text{and} \qquad
        \btY^{(d, \text{post})} = \sum_{i=1}^{r^{(d, \text{post})}} s_i \mu_i \nu_i^T,
    \end{align}
    where $r^{(d, \text{pre})}$ and $r^{(d,\text{post})}$ are hyper-parameters. 

    \item {\em Pre-Intervention - Synthetic Target Unit Construction}
    %
    : for every intervention $d \neq d^*$, define
    \begin{align} \label{eq:model_learning}
        \hlambda^{(d)} = \underset{w \in \Rb^{N^{(d)}}}{\text{argmin}} \sum_{t=1}^{T_0} \Big( y^{(1)}_{1t} - \sum_{n \in \Ic^{(d)}} w_n \cdot \ty^{(1)}_{nt} \Big)^2.
    \end{align}
    
    \item {\em Post-intervention Counterfactual Predictions}
    %
    : for every $t > T_0$,
    \begin{align} \label{eq:predictions}
        \hy^{(d)}_{1t} = \sum_{n \in \Ic^{(d)}} \hlambda^{(d)}_n \cdot \ty^{(d)}_{nt}. 
    \end{align}
\end{enumerate}

\subsubsection{Algorithm -- Mathematical Intuition}
Below we give intuition for the two major steps of the Synthetic Interventions algorithm (described above). 

\paragraph{Pre-Intervention - Synthetic Target Unit Construction.}
In order to predict the counterfactual outcome variables for our target unit under intervention $d$, we first build a synthetic version of our target unit using the donor units within partition $\Ic^{(d)}$.
That is, we find the set of weights, defined by $\hlambda^{(d)}$ in \eqref{eq:model_learning}, that best approximates the outcome variables of our target unit in the absence of any intervention. 
Crucially, to ensure that we are not over-fitting the data beyond the inherent model complexity in the target unit and donor trajectories, the Synthetic Interventions method performs Singular Value Thresholding (SVT) (equivalently Principal Component Analysis) on $\bY^{(d, \text{pre})}$ and $\bY^{(d, \text{post})}$ before learning the linear weights $\hlambda^{(d)}$, as seen in \eqref{eq:denoise}.
In other words, instead of simply performing Ordinary Least Squares (OLS) regression, we perform Principal Component Regression (PCR), which learns a model in the reduced subspace spanned by the columns of $\btY^{(d, \text{pre})}$, to ensure our model is appropriately de-noised and regularized.
For more motivation on why we use PCR, please refer to \cite{robust_pcr}.

\paragraph{Post-Intervention - Counterfactual Prediction.}
Once the model for intervention $d$ is learned, we re-scale the observed outcome variables associated with our donor units in $\Ic^{(d)}$ during the post-intervention period by the set of weights $\hlambda^{(d)}$ according to \eqref{eq:predictions}. Effectively, \eqref{eq:predictions} performs a synthetic intervention $d$ for our target unit and provides the corresponding counterfactual outcome variables under such a setting.
The key assumption that allows one to make valid counterfactual estimates using the Synthetic Interventions algorithm is that the relationship between units is invariant across interventions. 
Although this assumption may seem ``unnatural'', it is pleasingly implied (and justified) by a tensor factor model imposed on the data, which can be thought of as a natural extension to the matrix factor model generally assumed in panel data in econometrics (and, in particular, synthetic control methods). 
Mathematical details on this are beyond the scope of this memo, and an interested reader should feel free to contact the authors. 
Lastly, just as in the pre-intervention phase, we perform SVT (i.e., PCA) on the donor post-intervention data in \eqref{eq:denoise} for de-noising and regularization purposes.



\section{Acknowledgements}
We acknowledge fruitful discussions with Rahul Panicker of Wadhwani AI in helping us focus on mobility as the lens to take with respect to interventions.
We also thank Alberto Abadie of MIT Economics for helpful discussions.

\begin{appendix} 
\section{Looking Back: What Could Have Happened?} \label{sec:fig_pics1} 

Here, we display the figures described in Section \ref{sec:looking_back}. 

\begin{figure}[ht]
	\begin{subfigure}{0.5\textwidth}
		\centering
		\includegraphics[width=0.75\linewidth]{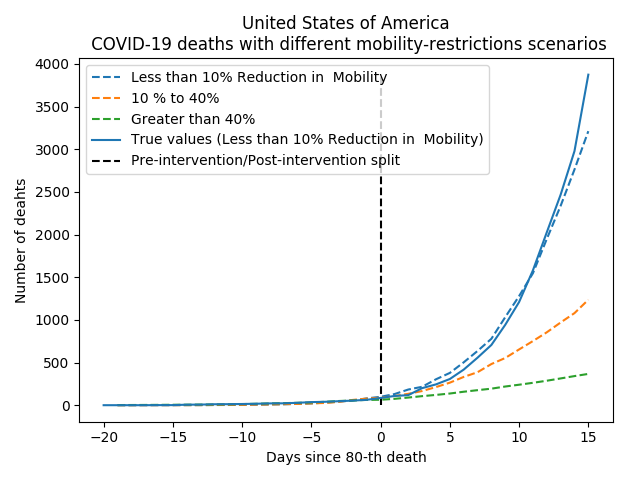}
		\caption{U.S. under all interventions.}
		\label{fig:us}
	\end{subfigure}
	~
	\begin{subfigure}{0.5\textwidth}
		\centering
		\includegraphics[width=0.75\linewidth]{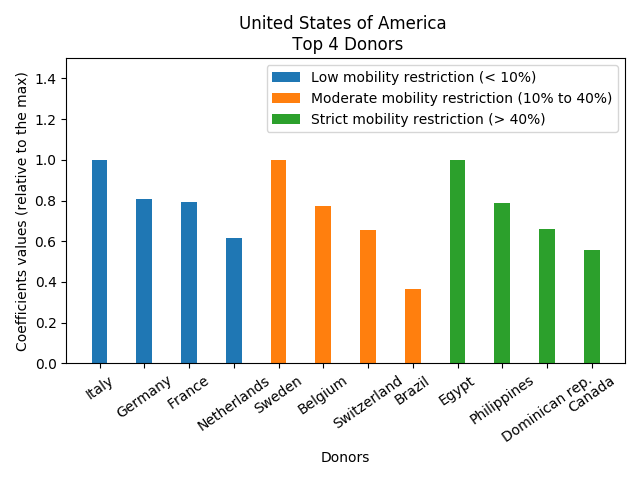}
		\caption{Top donor nations for the U.S.}
		\label{fig:donor_us}
	\end{subfigure} 
	~
	\begin{subfigure}{0.5\textwidth}
		\centering
		\includegraphics[width=0.75\linewidth]{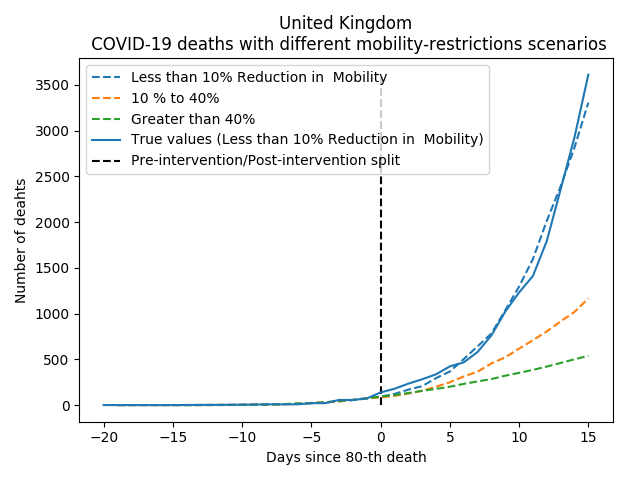}
		\caption{U.K. under all interventions.}
		\label{fig:uk}
	\end{subfigure}
	~
	\begin{subfigure}{0.5\textwidth}
		\centering
		\includegraphics[width=0.75\linewidth]{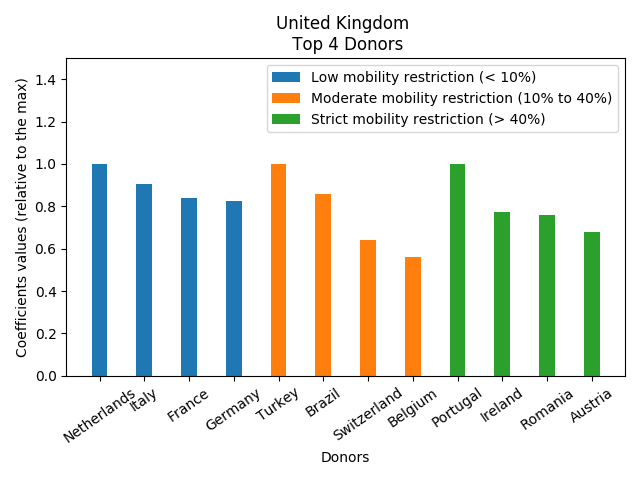}
		\caption{Top donor nations for the U.K.}
		\label{fig:donor_uk}
	\end{subfigure} 
	\caption{Validating Synthetic Interventions: countries with low mobility restricting interventions. }
	\label{fig:low_mobility}
\end{figure}

\begin{figure}[ht]
	\begin{subfigure}{0.5\textwidth}
		\centering
		\includegraphics[width=0.75\linewidth]{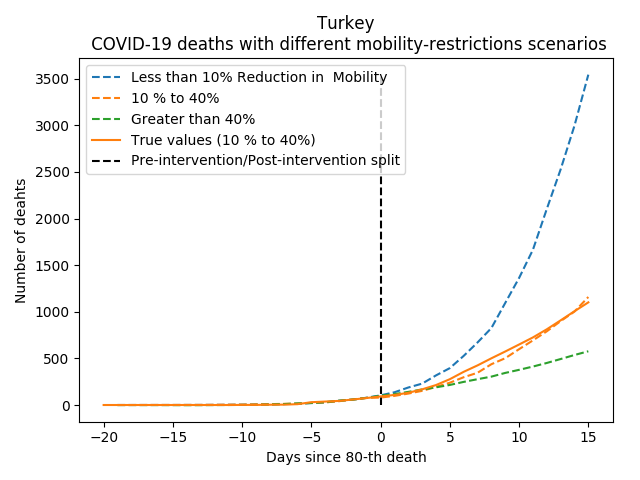}
		\caption{Turkey under all interventions.}
		\label{fig:turkey}
	\end{subfigure}
	~
	\begin{subfigure}{0.5\textwidth}
		\centering
		\includegraphics[width=0.75\linewidth]{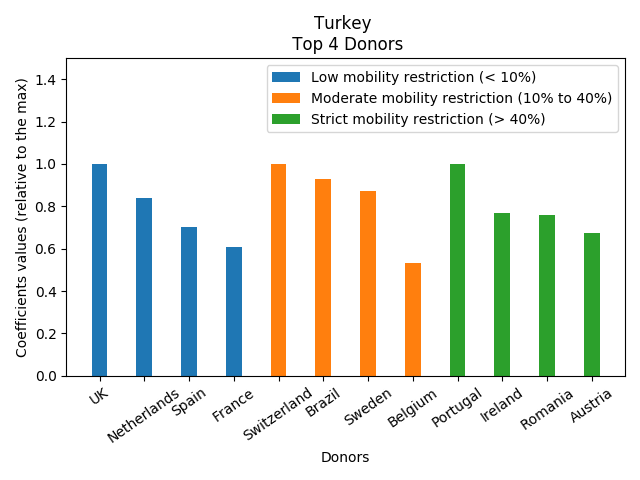}
		\caption{Top donor nations for Turkey.}
		\label{fig:donor_turkey}
	\end{subfigure} 
	~
	\begin{subfigure}{0.5\textwidth}
		\centering
		\includegraphics[width=0.75\linewidth]{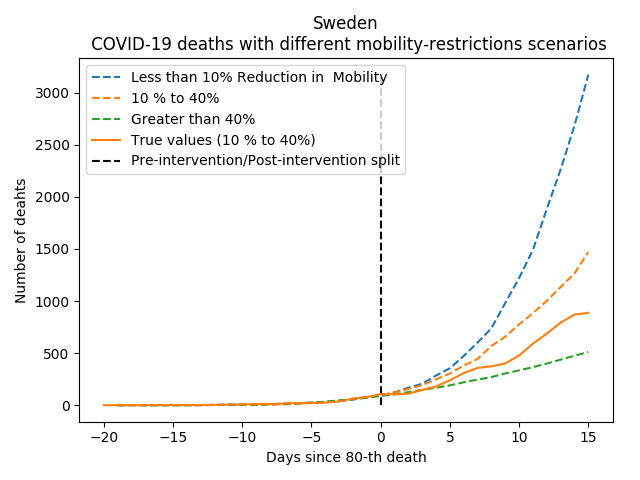}
		\caption{Sweden under all interventions.}
		\label{fig:sweden}
	\end{subfigure}
	~
	\begin{subfigure}{0.5\textwidth}
		\centering
		\includegraphics[width=0.75\linewidth]{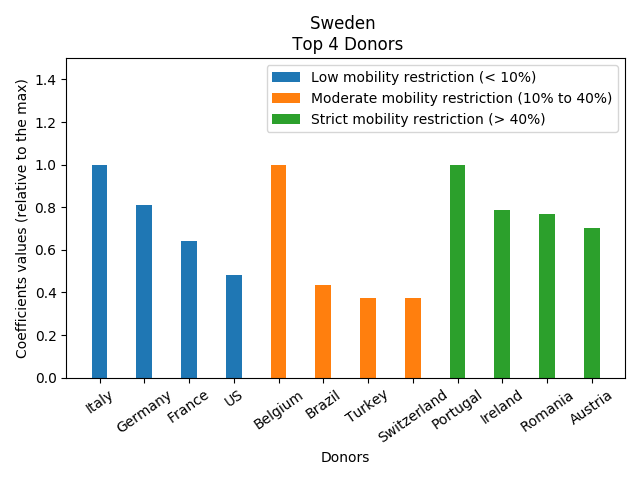}
		\caption{Top donor nations for Sweden.}
		\label{fig:donor_sweden}
	\end{subfigure} 
	\caption{Validating Synthetic Interventions: countries with moderate mobility restricting interventions. }
	\label{fig:moderate_mobility}
\end{figure}
~

\begin{figure}[H]
	\begin{subfigure}{0.5\textwidth}
		\centering
		\includegraphics[width=0.75\linewidth]{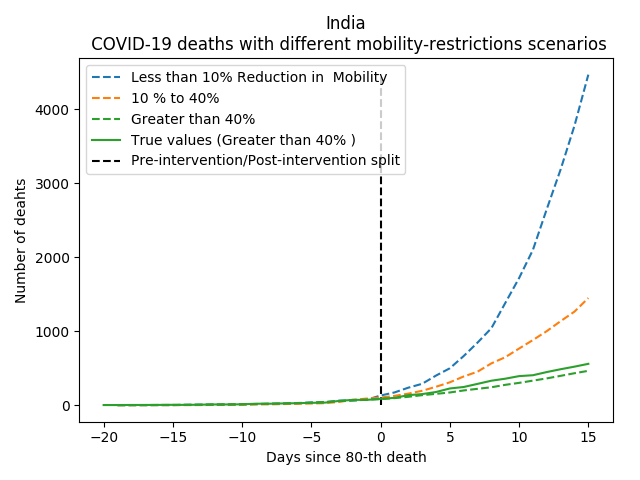}
		\caption{India under all interventions.}
		\label{fig:india}
	\end{subfigure}
	~
	\begin{subfigure}{0.5\textwidth}
		\centering
		\includegraphics[width=0.75\linewidth]{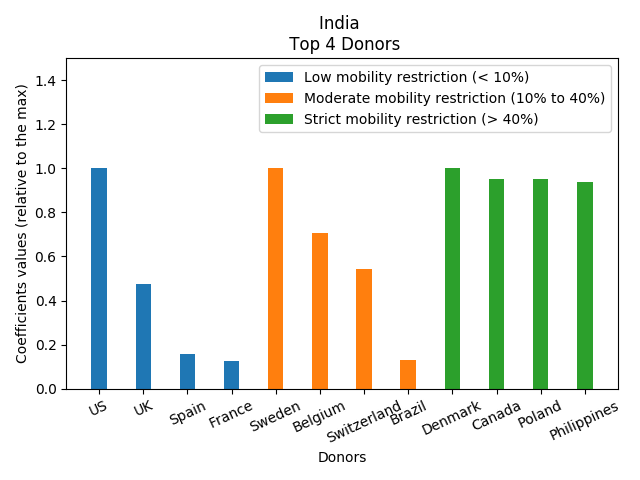}
		\caption{Top donor nations for India.}
		\label{fig:donor_india}
	\end{subfigure} 
	~
	\begin{subfigure}{0.5\textwidth}
		\centering
		\includegraphics[width=0.75\linewidth]{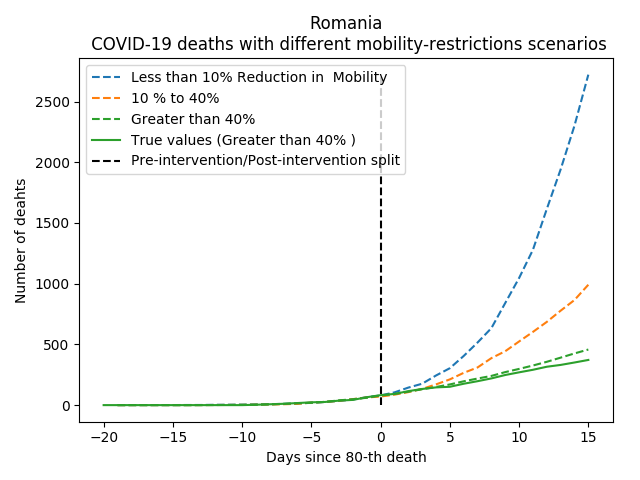}
		\caption{Romania under all interventions.}
		\label{fig:romania}
	\end{subfigure}
	~
	\begin{subfigure}{0.5\textwidth}
		\centering
		\includegraphics[width=0.75\linewidth]{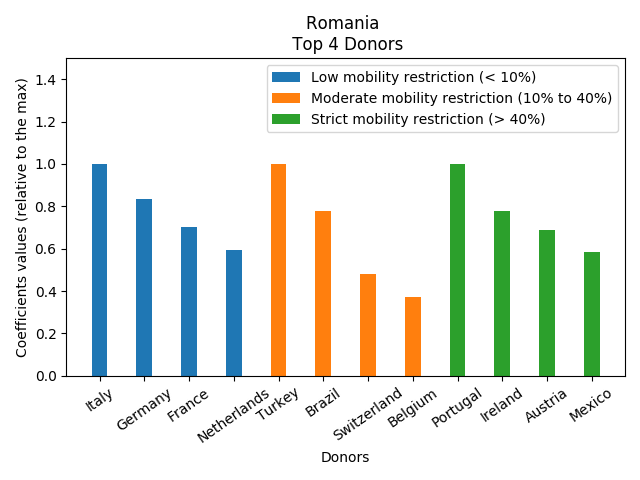}
		\caption{Top donor nations for Romania.}
		\label{fig:donor_romania}
	\end{subfigure} 
	\caption{Validating Synthetic Interventions: countries with severe mobility restricting interventions. }
	\label{fig:high_mobility}
\end{figure}

\newpage
\section{Looking Ahead: What Can Happen?} \label{sec:fig_pics2} 

Here, we display the figures described in Section \ref{sec:looking_forward}

\begin{figure}[H]
	\begin{subfigure}{0.9\textwidth}
		\centering
		\includegraphics[width=\linewidth]{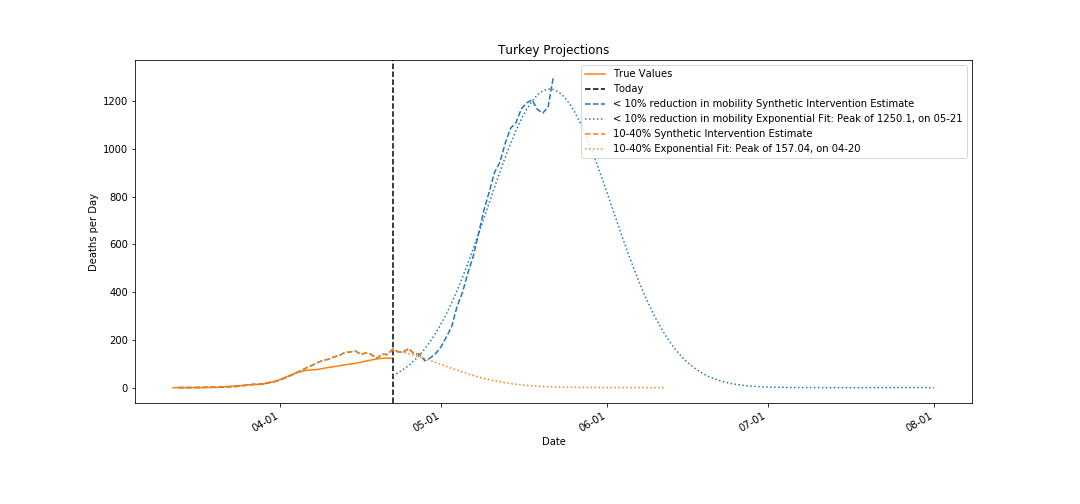}
		\caption{Projection for Turkey.}
		\label{fig:turkey_projections}
	\end{subfigure}
	~
	\begin{subfigure}{0.9\textwidth}
		\centering
		\includegraphics[width=\linewidth]{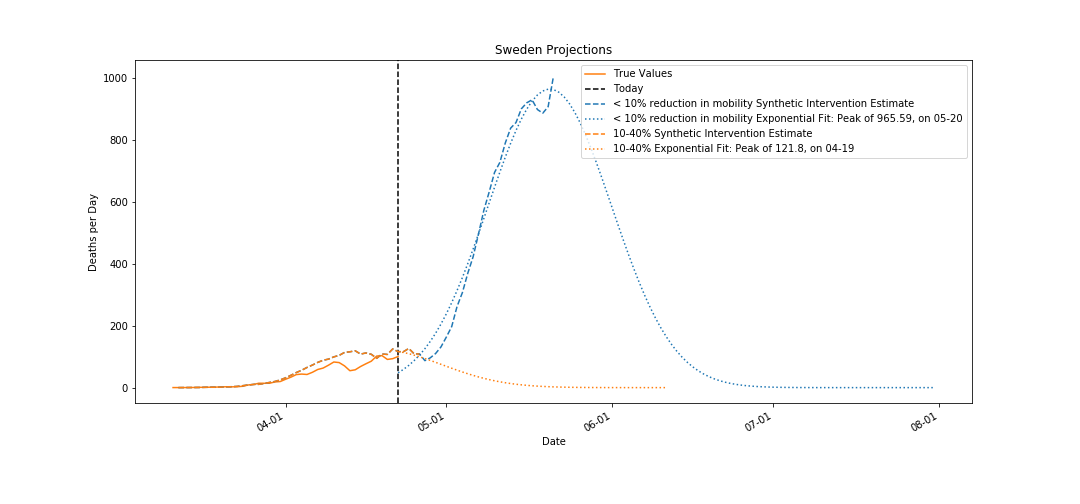}
		\caption{Projection for Sweden.}
		\label{fig:sweden_projections}
	\end{subfigure} 
	\caption{Looking ahead: projections for countries with moderate mobility restricting interventions.}
	\label{fig:projection_moderate}
\end{figure}

\newpage
\begin{figure}[H]
	\begin{subfigure}{0.9\textwidth}
		\centering
		\includegraphics[width=\linewidth]{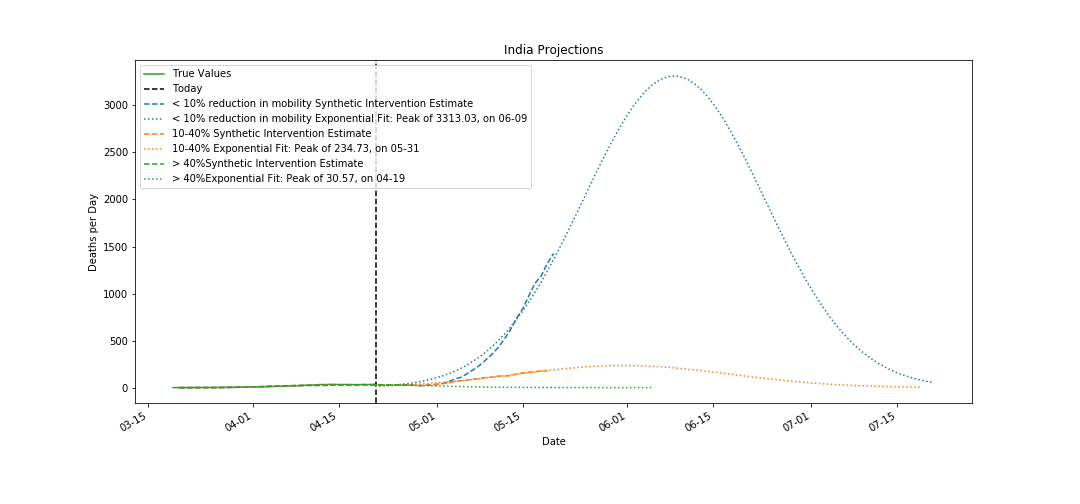}
		\caption{Projection for India.}
		\label{fig:india_projections}
	\end{subfigure}
	~
	\begin{subfigure}{0.9\textwidth}
		\centering
		\includegraphics[width=\linewidth]{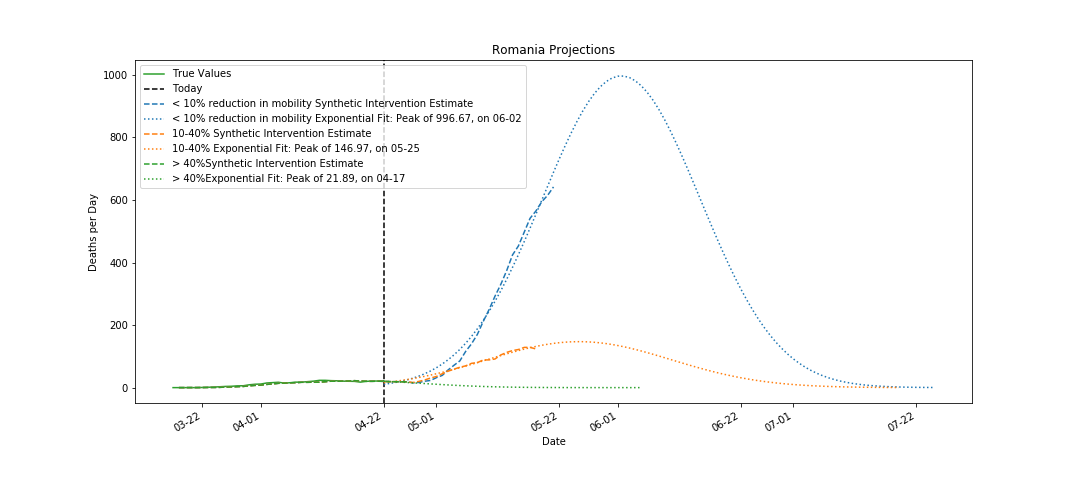}
		\caption{Projection for Romania.}
		\label{fig:romania_projections}
	\end{subfigure} 
	\caption{Looking ahead: projections for countries with severe mobility restricting interventions.}
	\label{fig:projection_severe}
\end{figure}

%

\vfill 
\end{appendix} 

\newpage
{\small
\bibliographystyle{abbrv}
\bibliography{bibliography} 

\begin{thebibliography}{1}

\bibitem{google_mobility}
{Google LLC} google covid-19 community mobility reports.
\newblock \url{https://www.google.com/covid19/mobility/}.
\newblock Accessed: 2020-04-20.

\bibitem{abadie_survey}
A.~Abadie.
\newblock Using synthetic controls: Feasibility, data requirements, and
  methodological aspects (working paper).
\newblock 2019.

\bibitem{SI}
A.~Agarwal, R.~Cosson, D.~Shah, and D.~Shen.
\newblock Synthetic interventions.
\newblock {\em CausalML Workshop of Advances of Neural Information Processing
  Systems (NeurIPS)}, 2019.

\bibitem{robust_pcr}
A.~Agarwal, D.~Shah, D.~Shen, and D.~Song.
\newblock On robustness of principal component regression.
\newblock {\em Advances of Neural Information Processing Systems (NeurIPS)},
  2019.

\bibitem{lag_study}
T.~B. L. B.~M. Wilson~N, Kvalsvig~A.
\newblock Case-fatality estimates for covid-19 calculated by using a lag time
  for fatality.
\newblock {\em Emerg Infect Dis.}, 2020.

\end{thebibliography}
}

\end{document}